\documentclass{article}
\DeclareUnicodeCharacter{02BB}{\textquoteleft}
\usepackage{microtype}
\usepackage{graphicx}
\usepackage{subfigure}
\usepackage{booktabs} 
\usepackage{multirow}

\usepackage{hyperref}

\usepackage[accepted]{icml2025}

\usepackage{amsmath}
\usepackage{amssymb}
\usepackage{mathtools}
\usepackage{amsthm}
\usepackage{colortbl}
\usepackage[capitalize,noabbrev]{cleveref}

\theoremstyle{plain}

\theoremstyle{definition}

\theoremstyle{remark}

\usepackage[textsize=tiny]{todonotes}

\icmltitlerunning{FedSCA: Federated Tuning with Similarity-guided Collaborative Aggregation for Heterogeneous Medical Image Segmentation}

\begin{document}

\twocolumn[
\icmltitle{FedSCA: Federated Tuning with Similarity-guided Collaborative Aggregation for Heterogeneous Medical Image Segmentation}



\icmlsetsymbol{equal}{*}

\begin{icmlauthorlist}
\icmlauthor{Yumin Zhang}{equal,yyy}
\icmlauthor{Yan Gao}{equal,comp,aaa}
\icmlauthor{Haoran Duan}{yyy}
\icmlauthor{Hanqing Guo}{sch}
\icmlauthor{Tejal Shah}{yyy}
\icmlauthor{Rajiv Ranjan}{yyy}
\icmlauthor{Bo Wei}{yyy}
\end{icmlauthorlist}

\icmlaffiliation{yyy}{School of Computing, Necastle University, UK}
\icmlaffiliation{comp}{Department of Computer Science and Technology, University of Cambridge, UK}
\icmlaffiliation{sch}{Department of Electrical and Computer Engineering, University of Hawaiʻi at Mānoa}
\icmlaffiliation{aaa}{Flower Labs}

\icmlcorrespondingauthor{Yumin Zhang}{y.zhang361@newcastle.ac.uk}

\icmlkeywords{Machine Learning, ICML}

\vskip 0.3in
]



\printAffiliationsAndNotice{\icmlEqualContribution} 

\begin{abstract}
Transformer-based foundation models (FMs) have recently demonstrated remarkable performance in medical image segmentation.
However, scaling these models is challenging due to the limited size of medical image datasets within isolated hospitals, where data centralization is restricted due to privacy concerns. These constraints, combined with the data-intensive nature of FMs, hinder their broader application.
Integrating federated learning (FL) with foundation models (FLFM) fine-tuning offers a potential solution to these challenges by enabling collaborative model training without data sharing, thus allowing FMs to take advantage of a diverse pool of sensitive medical image data across hospitals/clients.
However, non-independent and identically distributed (non-IID) data among clients, paired with computational and communication constraints in federated environments, presents an additional challenge that limits further performance improvements and remains inadequately addressed in existing studies.
In this work, we propose a novel FLFM fine-tuning framework, 
\underline{\textbf{Fed}}erated tuning with \underline{\textbf{S}}imilarity-guided \underline{\textbf{C}}ollaborative \underline{\textbf{A}}ggregation 
(FedSCA), encompassing all phases of the FL process. This includes (1) specially designed parameter-efficient fine-tuning (PEFT) for local client training to enhance computational efficiency; (2) partial low-level adapter transmission for communication efficiency; and (3) similarity-guided collaborative aggregation (SGCA) on the server side to address non-IID issues.
Extensive experiments on three FL benchmarks for medical image segmentation demonstrate the effectiveness of our proposed FedSCA, establishing new SOTA performance.
\end{abstract}

\section{Introduction}
\label{intro}

\begin{figure}
    \centering
    \includegraphics[width=1.0\linewidth]{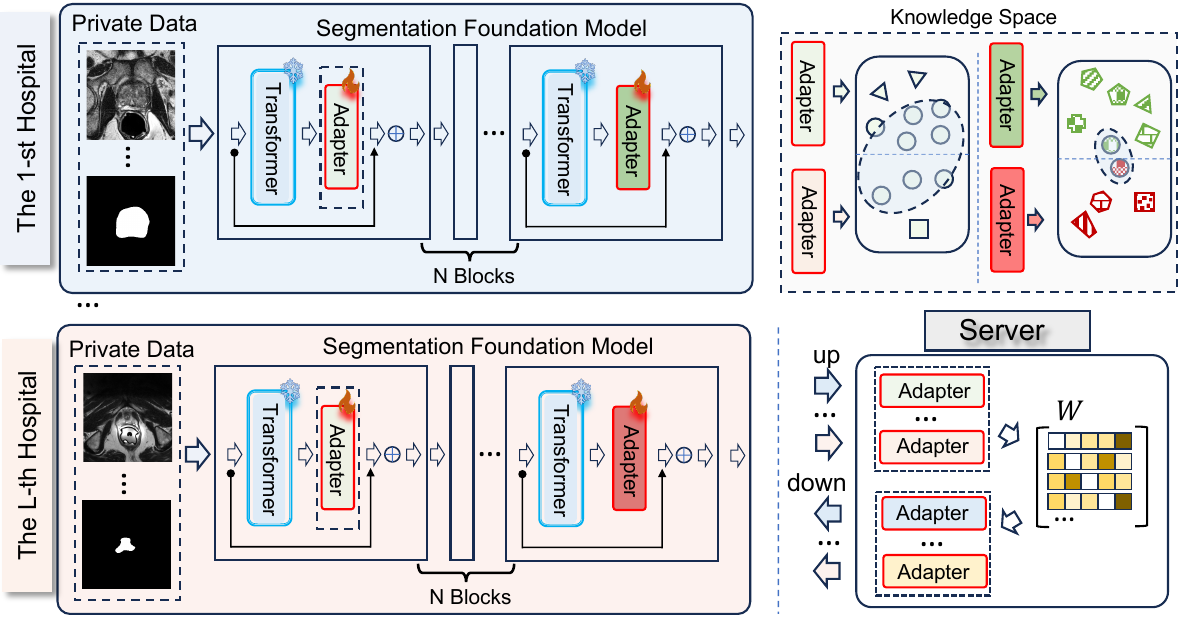}
    \caption{Illustration of the communication process during FL training. Compared to high-level client-specific semantic features, low-level model updates transmitted in the proposed FedECA contain more shared knowledge among clients.}
    \label{fig:fig1}
\end{figure}

Medical image segmentation~\cite{ronneberger2015u, zhang2024prototype} is critical for various medical applications, aiming to segment every pixel in a given medical image. 
Recently, the advent of transformer-based foundation models (FM) fine-tuning has inspired research~\cite{chen2021transunet, cao2022swin, ma2024segment} on applying these models to medical image segmentation tasks, achieving promising results.

However, compared to natural RGB images, medical images typically suffer from a significantly smaller sample size, which limits the effectiveness when using data-intensive foundation models. 
A straightforward approach to alleviate this limitation is aggregating medical data from multiple sources to build larger datasets.
Yet, due to the high sensitivity of medical data, this solution is often impractical as it contravenes privacy protection policies~\cite{rieke2020future}.
In this context, federated learning (FL)~\cite{mcmahan2017communication,gao2022end,rehman2022federated} offers a promising approach to distributed training by allowing collaborative model development without the need to share raw data. This enables secure access to valuable, sensitive medical data while safeguarding patient privacy~\cite{liu2021feddg,feng2022specificity, liu2024fedfms}.


The fast development of FM offers an advantageous starting point for federated model fine-tuning, facilitating faster convergence and enhancing overall performance~\cite{zhuang2023foundation,cai2023efficient, zhao2023fedprompt, zhao2024breaking}.
Despite advancements in FL for FM (\textit{i.e.}, FLFM), research on its application in medical image segmentation remains limited. 
Only a few studies have applied FLFM to medical image segmentation~\cite{liu2024fedfms,xu2022closing}, and their applicability is limited by the non-independent and identically distributed (non-IID) data across clients/hospitals.
Additionally, none of these existing studies have effectively addressed the pressing challenge posed by high communication costs in real-world medical environments.


To tackle these challenges, we propose a novel FLFM fine-tuning framework, \underline{\textbf{Fed}}erated tuning with \underline{\textbf{S}}imilarity-guided \underline{\textbf{C}}ollaborative \underline{\textbf{A}}ggregation (FedSCA), which incorporates three key components: (1) specially designed parameter-efficient fine-tuning (PEFT) techniques for local training on the client side, (2) partial low-level adapter transmission during the communication phase, and (3) similarity-guided collaborative aggregation (SGCA) on the server side. Specifically, we leverage PEFT techniques~\cite{houlsby2019parameter, pfeiffer2020adapterfusion} to facilitate reliable segmentation foundation models under limited training resources. During local training, only the parameters of adapter layers are optimized to enable and accelerate model training in clients with limited computational resources, while the remaining large parameters are kept frozen. 
Second, in FM fine-tuning, the lower-level layers (those closer to the input) typically capture shared representational knowledge, which can be leveraged effectively for aggregation in FL~\cite{jin2024exploring}. 
In contrast, higher-level layers (closer to the output) contain more client-specific information, which should remain local to preserve individual utility.
Guided by this insight, the framework only makes the low-level adapters be aggregated across clients, with reduced communication costs as a secondary advantage (Figure~\ref{fig:fig1}).
Third, to alleviate the non-IID issues, we introduce SGCA, an aggregation strategy based on the principle that clients with similar dataset distributions benefit most from mutual optimization directions. 
SGCA employs a collaborative relationship matrix to dynamically direct weight allocation during aggregation, fostering collaboration among similar clients with iterative updates at each communication round.
We summarize the main contributions as follows:

\begin{itemize}
    \item We propose a novel FLFM framework fine-tuning for medical image segmentation, termed FedSCA, which achieves robust performance on non-IID medical image segmentation tasks while operating within constrained computational resources.
    
    \item As the core element of the proposed FedSCA, SGCA  adaptively allocates aggregation weights to each client, promoting mutual learning among similar clients and mitigating the non-IID challenges commonly encountered in real-world federated environments.

    \item The partial low-level adapter transmission strategy facilitates the aggregation of shared knowledge among clients while reducing communication costs by \textbf{36x} compared to the baseline. 
    
    \item Extensive experimental results on three standard FL benchmarks for medical image segmentation illustrate the superiority of our proposed method, achieving up to an \textbf{2.0}\% increase in IoU and \textbf{1.9}\% in Dice, establishing new state-of-the-art (SOTA) results.
\end{itemize}

\section{Related Work}
\label{sec:relatedwork}

\subsection{Medical Image Segmentaion}
Medical image segmentation is a core task in medical image analysis. Among convolution-based models, U-Net~\cite{ronneberger2015u} and its successors~\cite{milletari2016v, oktay2018attention, isensee2021nnu, huang2020unet} have achieved notable success across a range of medical image segmentation applications. 
Recently, with the rise of foundation models~\cite{he2022masked, carion2020end, kirillov2023segment}, various works have explored how to leverage foundation models to implement medical image analysis.
For instance, RETFound~\cite{zhou2023foundation} utilizes large collections of unlabeled retinal images for pre-training a foundation model via self-supervised learning, which can then be efficiently adapted for disease detection tasks. 
Additionally, by incorporating CLIP embedding~\cite{radford2021learning}, the CLIP-driven universal model addresses label scarcity and limited dataset issues in organ segmentation and tumor detection tasks.
Furthermore, some works~\cite{wu2024medsegdiff, wu2024medsegdiff2} reformulate medical image segmentation as a mask generation task by adapting diffusion models~\cite{ho2020denoising}.
However, in practical settings, medical datasets are often distributed across various sites, and privacy-preserving policies restrict the collection of large datasets necessary for training medical foundation models.

\subsection{Federated Foundation Model}
Driven by the potential mutual benefits of increasing data availability and achieving more efficient convergence, the integration of federated learning (FL) with foundation models (FM) has gained substantial research interest~\cite{hilmkil2021scaling, kuang2024federatedscope}.
For instance, FedPrompt~\cite{zhao2023fedprompt} employs prompt tuning within an FL setting, significantly reducing both global communication and local training costs.  
Similarly, multilingual scenarios have been explored in~\cite{zhao2024breaking} by aggregating lightweight prompts.
AdaFL~\cite{cai2023efficient} incorporates adapters that are learnable only during local updates and used solely in communication phases, promoting faster convergence. 
Despite these advancements, relatively few studies have applied FLFM to medical image segmentation tasks due to limited computational resources and heterogeneous medical images.
One related work is FedFMS~\cite{liu2024fedfms}, which utilizes the Medical-SAM-Adapter~\cite{wu2023medical} within an FL framework.
However, FedFMS only implements the basic average aggregation method (\textit{i.e.}, FedAvg~\cite{mcmahan2017communication}), neglecting collaborative relationships among clients, thereby limiting performance improvements and resulting in significant degradation under non-IID data conditions. 

\section{Proposed Method}
\begin{figure*}[!t]
    
    \centering
    \includegraphics[width=1.0\linewidth]{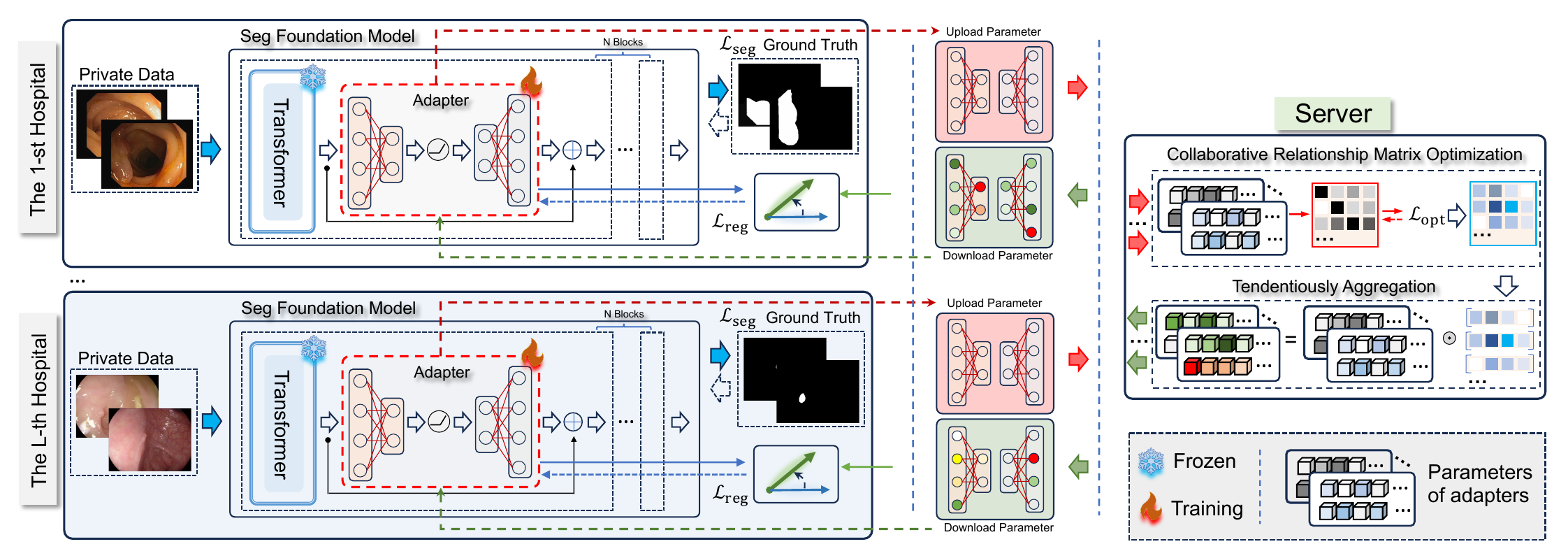}
    \caption{Pipeline of FedSCA, which mainly consists of three phases: federated parameter-efficient learning (FEL), low-level adapters transmission (LAT), and similarity-guided collaborative aggregation (SGCA) in one complete FL round.
    FEL: In the local training phase, only the parameters of inserted adapters are learnable.
    LAT: After local training, each client sends the low-lever adapter parameters to the server to share the universal representation knowledge among clients. 
    SGCA: Based on the collaborative relationship matrix, the server tendentiously aggregates these uploaded parameters to encourage similar clients to learn more from each other.}
    \label{fig:fig2}
\end{figure*}

\subsection{Problem Statement}

Following prior work on federated semantic segmentation setup in medical imaging~\cite{wang2022personalizing, liu2021feddg}, we assume a total of $N$ clients, ${\{\mathcal{C}_{i}\}}_{i=1}^{N}$, each possessing a distinct segmentation dataset $D_{i}=\{\boldsymbol{x}_{i}^{o}, \boldsymbol{y}_{i}^{o}\}_{o=1}^{n_i}$ composed of $n_i$ data-label pairs sampled from distribution $P_{i}$, where $x$ and $y$ are samples and labels, respectively.
The whole FL optimization has $R$ rounds, with each round consisting of a local training phase and a global aggregation phase.
Particularly, in the $r$-th round, the segmentation foundation model $F_{i}^{(r)}$ deployed on $i$-th client is updated on its private dataset $D_{i}$ by minimizing the local segmentation loss $\mathcal{L}_{\rm{seg}}$.

Inspired by recent FLFM tuning works~\cite{cai2023efficient, liu2024fedfms}, we apply PEFT to adapt local foundation models within limited computational resources.
Specifically, the foundation model $F_i$ can be mainly decomposed into $K$ blocks $\{B_i^k\}_{k=1}^{K}$, and each block with small adapter layers $A_i^k$ inserted. 
During the local training process, only the adapter parameters $\{\Theta_{A_i^k}^{(r)}\}_{k=1}^{K}$ are optimized, while parameters of transformer-based modules remain frozen. 
In the communication phase, instead of aggregating all adapters $\{\Theta_{A_i^k}^{(r)}\}_{k=1}^{K}$, we aggregate only the low-level adapters $\{\Theta_{A_i^l}^{(r)}\}_{l=1}^{L} (L < K)$, and higher adapters that encode client-specific semantic information are preserved on local to capture the unique data characteristics. 
By the above operation which we named low-level adapter transmission, we enable local models to obtain better trade-off between collaborative advantages brought by FL aggregation and individual local client utility.

However, the non-IID medical data across clients poses challenges by disrupting the consistency of model updates across clients, impeding FL convergence.
To tackle this challenge, based on the idea that clients with more similar datasets benefit from mutual optimization direction, we develop a similarity-guided collaborative aggregation (SGCA) to tendentiously allocate weight during the aggregation process for each client. 
By constructing a collaborative relationship matrix $\mathcal{W}^{(r)}\in\mathbb{R}^{N\times N}$, where $\mathcal{W}^{(r)}_{i,j}$ represents the similarity strength between $i$-th client and $j$-th client in the $r$-th round. 
Through $\{\Theta_{A_i^k}^{(r+1)}\}_{k=1}^{L}\leftarrow \sum_{j}\mathcal{W}^{(r)}_{i,j}\{\Theta_{A_j^k}^{(r)}\}_{k=1}^{L}$, SGCA guides these similar clients encouraged by more from each other, and mitigate negative efforts caused by dissimilar clients, reaching a higher performance without high communication costs require.

\subsection{FedSCA}

\subsubsection{Overview}

The pipeline of our method is visualized in Figure~\ref{fig:fig2}, which mainly consists of federated parameter-efficient learning (FEL) and similarity-guided collaborative aggregation (SGCA) corresponding to local training and global aggregation two phases in each FL round, respectively.
Specifically, during the local training process, local foundation models are tuned in parallel via PEFT. 
After local training finished, each client send low $L$ layers $\{\Theta_{A_i^k}^{(r)}\}_{k=1}^{L}$ to participate in low-level adapters transmission (LAT).
On the server side, based on the proposed SGCA, we tendentiously allocate weight for each client to continue the training in the next round.
FEL, LAT, and SGCA are illustrated in detail in the following subsections.

\subsubsection{Federated Parameter-Efficient Learning}
Directly full-tuning the foundation models requires significant computational resources, making it impractical for many medical clients.
Inspired by the success of implementing PEFT techniques in FL~\cite{cai2023efficient, liu2024fedfms}, we leverage the adapter approaches~\cite{houlsby2019parameter, pfeiffer2020adapterfusion} that introduce compact adapter layers within transformer blocks, thereby lowering local training expenses.

An adapter typically consists of a down-projection layer $M_d \in \mathbb{R}^{u \times d}$ followed by a non-linear activation function $\sigma(\cdot)$, and completed by an up-projection layer $M_u \in \mathbb{R}^{d \times u}$, where $u$ and $d$ denote the corresponding dimensions.
This forward computation, incorporating a residual connection, can be denoted as:
\begin{equation}
    A(\boldsymbol{f}) = M_u \cdot \sigma(M_d \cdot \boldsymbol{f}) + \boldsymbol{f}.
\end{equation}
Here, $\boldsymbol{f}$ denotes the feature implemented by adapter $A(\cdot)$.
For the foundation model $F_{i}$ generally consists of $K$ blocks $B_i^{k}$ where: 
\begin{equation}
    F_{i}(\boldsymbol{x}_i; \Theta_{F_i}) = B_{i}^{K}(\cdots B_{i}^{1}(\boldsymbol{x}_{i};\Theta_{B_i^1})\cdots;\Theta_{B_i^{K}}).
\end{equation}
Here, $\Theta_{B_i^{k}}$ denotes the parameter of $k$-th block that can be further decomposed into the transformer module $T_{i}^{k}$ and adapter layer $A_{i}^{k}$ two parts: $\Theta_{B_i^k} = \{\Theta_{T_i^k}, \Theta_{A_i^k}\}$.
During the local training process, only $\Theta_{A_i^k}$ is learnable, and $\Theta_{T_i^k}$ is frozen. 
Consequently, the objective function for the $i$-th client in $r$-th round is expressed as:
\begin{equation}
    \begin{aligned}
    \mathcal{L}_{\mathrm{seg}}({\Theta_{{A_i}}^{(r)}})& = \mathbb{E}_{(\boldsymbol{x}_{i}^{o}, \boldsymbol{y}_{i}^{o})\in D_{i}}[ -\boldsymbol{y}_i^{o}\log {F_i}(\boldsymbol{x}_{i}^{o};\Theta_{F_i}^{(r)})\\ & - (1 - \boldsymbol{y}_{i}^{o})\log (1 - {F_i}(\boldsymbol{x}_{i}^{o};\Theta_{F_i}^{(r)})].
    \end{aligned}
    \label{eq:eq3}
\end{equation}
We use $\Theta_{A_i}^{(r)}=\{\Theta_{A_i^1}^{(r)}, \Theta_{A_i^2}^{(r)}, \cdots, \Theta_{A_i^K}^{(r)}\}$ to denote the parameters of all adapters within the foundation model $F_i^{(r)}$.
Given that $\sum_{k=1}^{K}|\Theta_{A_i^k}^{(r)}|\ll \sum_{k=1}^{K}|\Theta_{T_i^k}^{(r)}|$, the local training costs are dramatically reduced. 

\subsubsection{Low-level Adapter Transmission}

\begin{figure}
    \centering
    \includegraphics[width=1.0\linewidth]{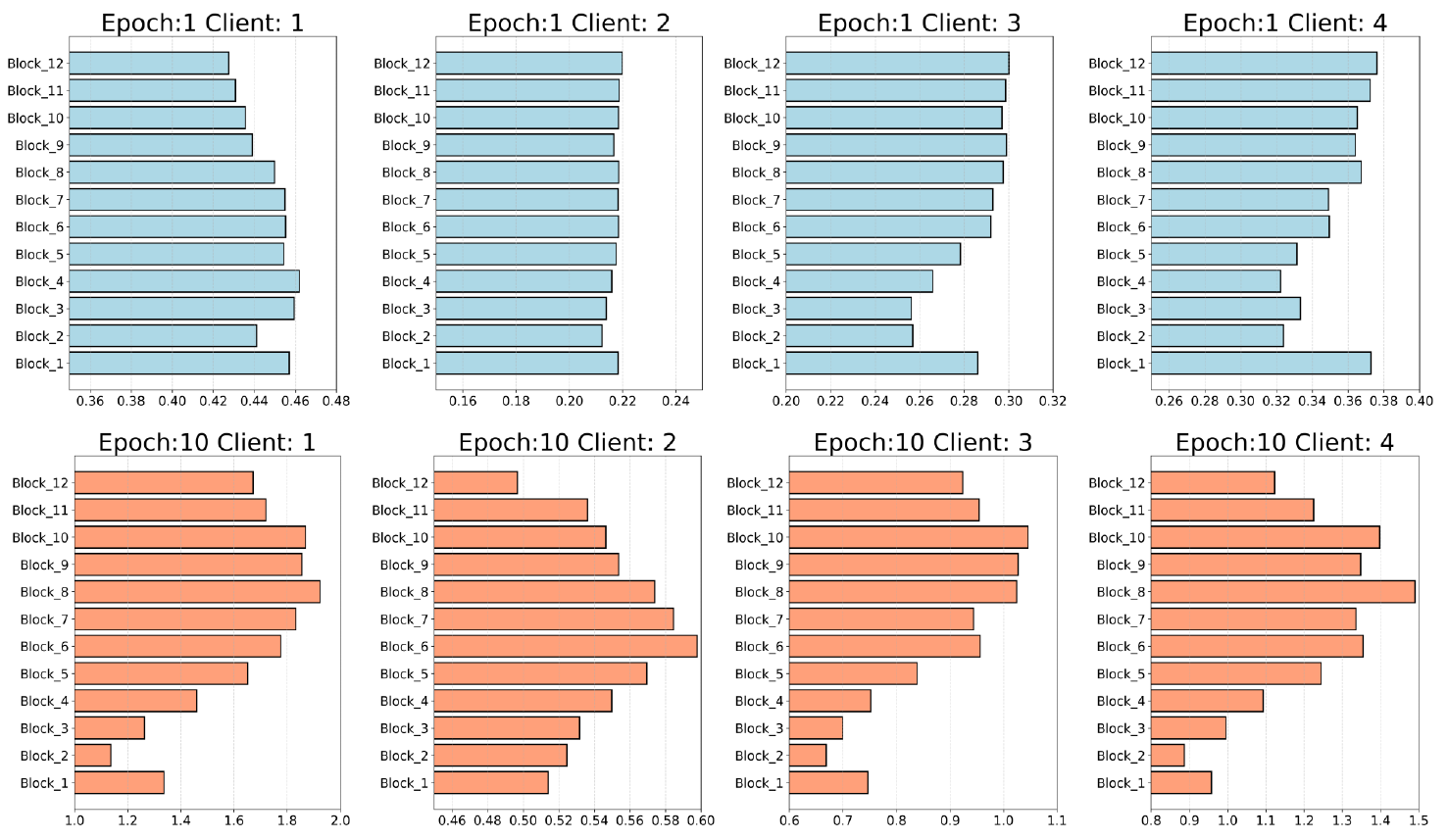}
    \caption{During local training (epoch 1 - 10) on the FedPolyp dataset, the parameters of the lower layers shift less than those of the higher layers.}
    \label{fig:parashift}
\end{figure}

In previous works~\cite{cai2023efficient,liu2024fedfms}, all learnable parameters of the local FM are communicated by default in the $r$-th round: 
\begin{equation}
    \Theta_{A}^{(r)} = \mathrm{Agg}(\Theta_{A_1}^{(r)},\Theta_{A_2}^{(r)},\cdots,\Theta_{A_N}^{(r)}).
    \label{eq:eq4}
\end{equation}
Here, $\mathrm{Agg}(\cdot)$ denotes the implemented aggregation algorithm.
However, not all learnable parameters are necessary for communication.
Typically, parameters with significant changes capture personalized knowledge, while those that remain largely unchanged should be shared across clients~\cite{tamirisa2024fedselect}.
Considering this, we first measure the parameters shift and visualize them in Figure~\ref{fig:parashift}.
Notably, the shift of parameters is not uniform, showing that with the training, low layers tend to be stable while high layers undergo more significant changes. 
Inspired by this, we decomposed the $\{\Theta_{A_i^k}^{(r)}\}_{k=1}^{K}$ into two parts: low $L$ adapters $\{\Theta_{A_i^l}^{(r)}\}_{l=1}^{L}$ and high $(K-L)$ adapters $\{\Theta_{A_i^h}\}_{h=L+1}^{K}$, and only the $\{\Theta_{A_i^l}^{(r)}\}_{l=1}^{L}$ are involved in communication.
Without loss of generality, we set $L=1$, and thus the Equation~(\ref{eq:eq4}) should be adapted as:
\begin{equation}
    \Theta_{A^{1}}^{(r)} = \mathrm{Agg}(\Theta_{A_1^1}^{(r)}, \Theta_{A_2^1}^{(r)}, \cdots, \Theta_{A_N^1}^{(r)}).
\end{equation}
However, the inherent non-IID characteristics, if not properly addressed, still hinder FL convergence and compromise the effectiveness of collaborative universal representation.

\subsubsection{Similarity-Guided Collaborative Aggregation}

Aiming to tackle the above challenge, we develop a novel aggregation method named similarity-guided collaborative aggregation (SGCA).
Specifically, we assume that the clients with similar dataset distributions should exhibit similar optimization directions and, therefore, benefit more from mutual learning~\cite{luo2021no,ye2023personalized}.
Let $\mathcal{W}^{N\times N}$ represent a collaborative relationship matrix, where $\mathcal{W}_{ij} \geq 0$ indicates the similarity strength between client $i$ and client $j$, satisfying the condition $\sum_{j=1}^{N}\mathcal{W}_{ij}=1$.
In a centralized training manner, client optimization with PEFT can be formulated as:
\begin{equation}
    \min_{\hat{\Theta}_{A_{i}}, \mathcal{W}}\sum_{i=1}^{N}w_{i} \left(\mathcal{L}_{\rm{seg}}(\hat{\Theta}_{A_i})- \alpha \sum_{j=1}^{N}\mathcal{W}_{ij}d(\Theta_{A_i},\Theta_{A_j})\right),
    \label{eq:eq6}
\end{equation}
where $\hat{\Theta}_{A_i} = \sum_{j=1}^{N}\mathcal{W}_{ij}\Theta_{A_j}$, and $\alpha$ is the hyperparameter.
We use $d(\cdot)$ to measure the similarity between parameter sets from different clients.
However, since the private datasets are retained locally, and parameters from different clients are only accessible on the server, $\hat{\Theta}_{A_{i}}$ and $\mathcal{W}$ in Equation~(\ref{eq:eq6}) cannot be solved simultaneously.

\begin{algorithm}[!t]
    \begin{algorithmic}[1]
    \caption{Optimization of FedSCA}
    \label{alg1}
    \REQUIRE $N$ initial local models $\{{F_i}\}_{i=1}^{N}$ with corresponding private dataset $\{D_{i}\}_{i=1}^{N}$, total communication round $R$, local learning rate $\eta$, collaborative relationship matrix $\mathcal{W}^{(0)}$
    \ENSURE Trained local models $\{F_1, F_2, \cdots F_N\}$ 
   
    \FOR{communication round $r=1, \cdots, R$}
    \STATE Server sends $\Theta_{A_i^1}^{(r)} = \sum_{j}\mathcal{W}_{ij}^{(r-1)}\Theta_{A_i^1}^{(r-1)}$ to clients
    \FOR{client $i=1, \cdots, N$ parallel}
    \STATE Optimize $\{\Theta_{A_i^k}\}_{k=1}^{K}$ by Equation~(\ref{eq:eq9}). 
     \texttt{\#FEL}
    \ENDFOR
    \STATE clients upload $\{\Theta_{A_i^1}\}_{i=1}^{N}$ to the server. \texttt{\#LAT}
    \STATE Optimize $\mathcal{W}^{(r)}$ by solving Equation~(\ref{eq:eq11}). \texttt{\#SGCA}
    \ENDFOR
    \end{algorithmic}
\end{algorithm}

Therefore, to solve the Equation~(\ref{eq:eq6}), we decomposed it into cooperative optimization steps: the collaborative relationship matrix $\mathcal{W}$ is updated on the server, while each client independently trains $\hat{\Theta}_{A_i}$ using its private dataset $D_i$.

\vspace{1mm}
\noindent
\textbf{SGCA on the Server.}
In the $r$-th FL round, the server can only access the uploaded parameters $\{\Theta^{(r)}_{A_i^1}\}_{i=1}^{N}$, without additional information can be utilized.
We assume that clients with larger datasets are more reliable. 
For each client $i$, we optimize the collaborative relationship matrix via:
\begin{equation}
\begin{aligned}
\mathcal{L}_{\rm{opt}}(\mathcal{W}^{(r)}_{ij}) = 
    \sum_{j=1}^{N}(\mathcal{W}^{(r)}_{ij}- m_{i})^2 - \alpha \sum_{j=1}^{N}\mathcal{W}^{(r)}_{ij}d(\Theta^{(r)}_{A_i^1}, \Theta^{(r)}_{A_j^1}),
\end{aligned}
\label{eq:eq11}
\end{equation}
where $m_i = n_i / \sum_{j} n_j$ represents the relative weight based on dataset size, $\alpha$ is a hyperparameter carried over from Equation~(\ref{eq:eq6}), and the second term promotes the $\mathcal{W}^{(r)}$ to allocate greater attention to these similar parameters. 
In practice, we implement $d(\cdot)$ using the $L_2$ norm distance and rewrite Equation~(\ref{eq:eq11}) as a standard quadratic program, which can be solved using conventional convex optimization solvers~\cite{diamond2016cvxpy, agrawal2018rewriting}.

\vspace{1mm}
\noindent
\textbf{FEL on the Clients.}
After computing the collaborative relationship matrix, each client receives the corresponding parameters as $\Theta_{A_i^1}^{(r+1)} \leftarrow \sum_{j=1}^{N}\mathcal{W}_{ij}^{(r)}\Theta_{A_j^1}^{(r)}$.
At the beginning of the $(r+1)$-th FL round, the learnable parameters for $i$-th client can be summarized as: 
\begin{equation}
    \Theta_{A_i}^{(r+1)} = \{\sum_{j=1}^{N}\mathcal{W}_{ij}^{(r)}\Theta_{A_j^1}^{(r)}, \Theta_{A_i^2}^{(r)}, \cdots, \Theta_{A_i^K}^{(r)}\}.
\end{equation}
Then, the model $F_i$ deployed on the $i$-th client is optimized by minimizing the objective function $\mathcal{L}_{\rm{obj}}$, defined as:
\begin{equation}
    \mathcal{L}_{\rm{obj}}(\Theta_{A_i}^{(r+1)}) = \mathcal{L}_{\rm{seg}}(\Theta_{A_i}^{(r+1)}) + \beta \mathcal{L}_{\rm{reg}} (\Theta_{A_i^1}^{(r+1)}).
    \label{eq:eq9}
\end{equation}
Here, $\beta$ is the hyperparameter, and we regularize the $\Theta_{A_i^1}^{(r+1)}$ by the regularization function to avoid the overfitting issue:
\begin{equation}
    \mathcal{L}_{\rm{reg}}(\Theta_{A_i^1}^{(r+1)}) = -\cos{(\Theta_{A_i^1}^{(r+1)}, \sum_{j=1}^{N}\mathcal{W}_{ij}^{(r)}\Theta_{A_j^1}^{(r)})}.
    \label{eq:eq10}
\end{equation}
After completing local training for the $(r+1)$-th FL round, each client uploads $\{\Theta_{A_i^1}^{(r+1)}\}_{i=1}^{N}$ to the server, repeating the above process until the final FL round.
The optimization steps of FedSCA are summarized in Algorithm~\ref{alg1}.

\section{Experiments}

\begin{table*}[!t]
    \centering
    \caption{Segmentation results of our proposed method compared with SOTA on Fed-Prostate dataset. Two evaluation metrics are used — Dice and IoU (higher is better). The best results are highlighted in \textbf{bold}, while the second-best results are \underline{underlined}.}
    \begin{tabular}{l |  c c c c c c c | c c c c c c c }
    \toprule
    \multirow{2}{*}{\textbf{Method}} & \multicolumn{7}{c|}{Fed-Prostate (Dice\%)} & \multicolumn{7}{c}{Fed-Prostate (IoU\%)}\\
    \cline{2-15}
    & $\mathcal{C}_{1}$ &$\mathcal{C}_{2}$ & $\mathcal{C}_{3}$ & $\mathcal{C}_{4}$ & $\mathcal{C}_{5}$ & $\mathcal{C}_{6}$ & Avg. & $\mathcal{C}_{1}$ &$\mathcal{C}_{2}$ & $\mathcal{C}_{3}$ & $\mathcal{C}_{4}$ & $\mathcal{C}_{5}$ & $\mathcal{C}_{6}$ & Avg. \\
    \midrule
    FedProx$^{\dag}$ & 84.6 & 89.8 & 86.0  & 89.2 & 89.1 & 87.0 & 88.2 & 75.1 & 82.3 & 77.0 & 81.1 & 81.7 & 78.2 & 80.0 \\
    FedFMS$^{\dag}$ & 87.0 & 90.5 & 86.8 & 90.8 & 89.5 & 84.6 & 89.0 & 78.6 & 83.2 & 77.5 & 83.8 & 82.7 & 76.1 & 81.4 \\
    FedHEAL$^{\dag}$ & 87.8 & 90.6 & 88.0 & 90.5 & 90.5 & 83.7 & 89.3 & 79.8 & 83.5 & 79.2 & 83.3 & 83.7 & 74.5 & 81.8  \\
    FedDBE$^{\dag}$& 86.2 & 90.0 & 85.7 & 90.6 & 89.2 & 86.8 & 88.8 & 78.4 & 82.5 & 76.3 & 83.2 & 82.3 & 78.2 & 81.1 \\
    FedRA$^{\dag}$ & 85.3 & 88.6 & 86.7 & 89.7 & 87.6 & 85.2 & 87.7 & 76.3 & 80.2 & 77.8 & 81.9 & 79.7 & 75.7 & 79.4 \\
    FedSA$^{\dag}$ & 88.8 & 91.2 & \underline{88.3} & 90.4 & \underline{91.1} & \underline{88.9} & 90.2 & 81.1 & 84.5 & \underline{79.6} & 83.3 & \underline{84.2} & \underline{80.9} & 82.9  \\
    FedALA$^{\dag}$ & \underline{89.4} & \underline{91.6} & 86.9 & \underline{92.5} & 90.4 & 88.4 & \underline{90.5} & \underline{81.7} & \underline{85.0} & 78.4 & \underline{86.2} & 83.9 & 80.8 & \underline{83.6} \\
     \textbf{Ours} & \textbf{90.6} &	\textbf{91.7} &	\textbf{91.2} &	\textbf{93.0}	& \textbf{91.7} & \textbf{89.1} &	\textbf{91.6} & \textbf{83.4} & \textbf{85.3} & 	\textbf{84.2} &	\textbf{87.2} & \textbf{85.4} & 	\textbf{81.6} &	\textbf{85.1} \\
    \bottomrule
    \end{tabular}
    \label{tab:tab1}
\end{table*}

\begin{table*}[!t]
    \centering
    \small
    \caption{Segmentation results of our proposed method compared with SOTA on Fed-Polyp and Fed-Fundus datasets. Dice and IoU are used as evaluation metrics (higher is better). The best results are highlighted in \textbf{bold}, while the second-best results are \underline{underlined}.}
    \setlength{\tabcolsep}{1.1 mm}{
    \begin{tabular}{l |  c c c c c | c c c c c |  c c c c c | c c c c c}
    \toprule
    \multirow{2}{*}{\textbf{Method}} & \multicolumn{5}{c|}{Fed-Polyp (Dice\%)} & \multicolumn{5}{c|}{Fed-Polyp (IoU\%)} & \multicolumn{5}{c|}{Fed-Fundus (Dice\%)} & \multicolumn{5}{c}{Fed-Fundus (IoU\%)}\\
    \cline{2-21}
    & $\mathcal{C}_{1}$ &$\mathcal{C}_{2}$ & $\mathcal{C}_{3}$ & $\mathcal{C}_{4}$ &  Avg. & $\mathcal{C}_{1}$ &$\mathcal{C}_{2}$ & $\mathcal{C}_{3}$ & $\mathcal{C}_{4}$ &  Avg. & $\mathcal{C}_{1}$ &$\mathcal{C}_{2}$ & $\mathcal{C}_{3}$ & $\mathcal{C}_{4}$ &  Avg. & $\mathcal{C}_{1}$ &$\mathcal{C}_{2}$ & $\mathcal{C}_{3}$ & $\mathcal{C}_{4}$ &  Avg. \\

    \midrule
    FedProx$^{\dag}$ & 88.8 & 60.5 & 68.0 & 87.7 & 82.3 & 81.9 & 53.5 & 60.4 & 79.6 & 74.9 & 77.7 & 72.1 & 88.2 & 88.6 & 82.8 & 68.2 & 61.1 & 80.0 & 80.5 & 73.7 \\
    FedFMS$^{\dag}$ & 88.0 & 75.7 & 62.6 & 89.6 & 82.9 & 81.2 & 67.2 & 55.5 & 83.1 & 76.0 & 87.6 & 78.2 & 90.3 & 90.8 & 87.3 & 80.5 & 68.1 & 83.1 & 83.9 & 79.5 \\
    FedHEAL$^{\dag}$ & 88.0 & 77.2 & 67.1 & 90.7 & 84.2 & 81.0 & 68.9 & 61.0 & 84.2 & 77.3 & 87.4 & 82.2 & 90.5 & 90.8 & 88.2 & 79.7 & 72.5 & 83.3 & 84.0 & 80.4 \\
    FedDBE$^{\dag}$ & 87.6 & \underline{81.6} & \underline{69.7} & 91.1 & 84.9 & 80.5 & 73.2 & 62.2 & 84.6 & 77.8 & 89.3 & 69.6 & 91.0 & 90.7 & 85.9 & 82.7 & 59.7 & 84.1 & 83.6 & 78.3 \\
    
    FedRA$^{\dag}$ & 83.9 & 45.3 & 55.6 & 82.4 & 75.0 & 76.1 & 37.4 & 48.0 & 73.6 & 67.0 & 89.7 & 79.0 & 90.8 & 91.1 & 88.1 & 83.1 & 69.1 & 83.9 & 84.3 & 80.6 \\
    FedSA$^{\dag}$ & 89.2 & 72.4 & 68.3 & 88.5 & 83.8 & 82.6 & 64.9 & \underline{62.5} & 81.1 & 77.1 & \textbf{91.3} & 80.8 & 91.2 & 91.3 & 88.9 & \textbf{85.4} & 72.2 & 84.5 & 84.7 & 82.0 \\

    FedALA$^{\dag}$ & \underline{89.3} & 81.3 & 66.4 & \underline{91.6} & \underline{85.2} & \underline{83.3} & \underline{73.7} & 61.1 & \underline{85.8} & \underline{79.2} & \underline{90.7} & \textbf{86.8} & \underline{91.7} & \underline{91.7} & \underline{90.4} & 84.3 & \textbf{78.3} & \underline{85.3} & \underline{85.2} & \underline{83.5} \\
    \textbf{Ours} & \textbf{89.5}	& \textbf{82.7} &	\textbf{76.1}	& \textbf{91.9} &	\textbf{87.2} & \textbf{83.4}	& \textbf{76.3}	& \textbf{69.2} &	\textbf{86.1} & 	\textbf{81.1} & \textbf{91.3} & \underline{86.4} & 	\textbf{91.8} & 	\textbf{92.0} & 	\textbf{90.6} & \underline{85.0}	& \underline{77.8} & 	\textbf{85.6} & 	\textbf{85.6}	& \textbf{83.8} \\
    \bottomrule
    \end{tabular}}
    \label{tab:tab2}
\end{table*}

\subsection{Federated Datasets Settings}
Following previous works~\cite{chen2024think, liu2021feddg}, we conduct experiments on three federated medical image segmentation datasets: Fed-Prostate~\cite{liu2021feddg}, Fed-Polyp~\cite{wang2022personalizing}, and Fed-Fundus~\cite{liu2021feddg}:

\vspace{1mm}
\noindent
\textbf{Fed-Prostate}~\cite{liu2021feddg} is collected from six different clients for prostate MRI segmentation, and the sample size of each site is $\{261, 384, 158, 468, 421, 175\}$.

\vspace{1mm}
\noindent
\textbf{Fed-Polyp}~\cite{wang2022personalizing} is collected from four different data sources with the sample size of $\{1000, 196, 379, 612\}$, which is utilized for polyp segmentation.

\vspace{1mm}
\noindent
\textbf{Fed-Funds}~\cite{liu2021feddg} is for optic disc and cup segmentation, which is collected and annotated from four clients with the sample of $\{101, 159, 400, 400\}$.

\begin{figure}[!t]
    \centering
    \includegraphics[width=1\linewidth]{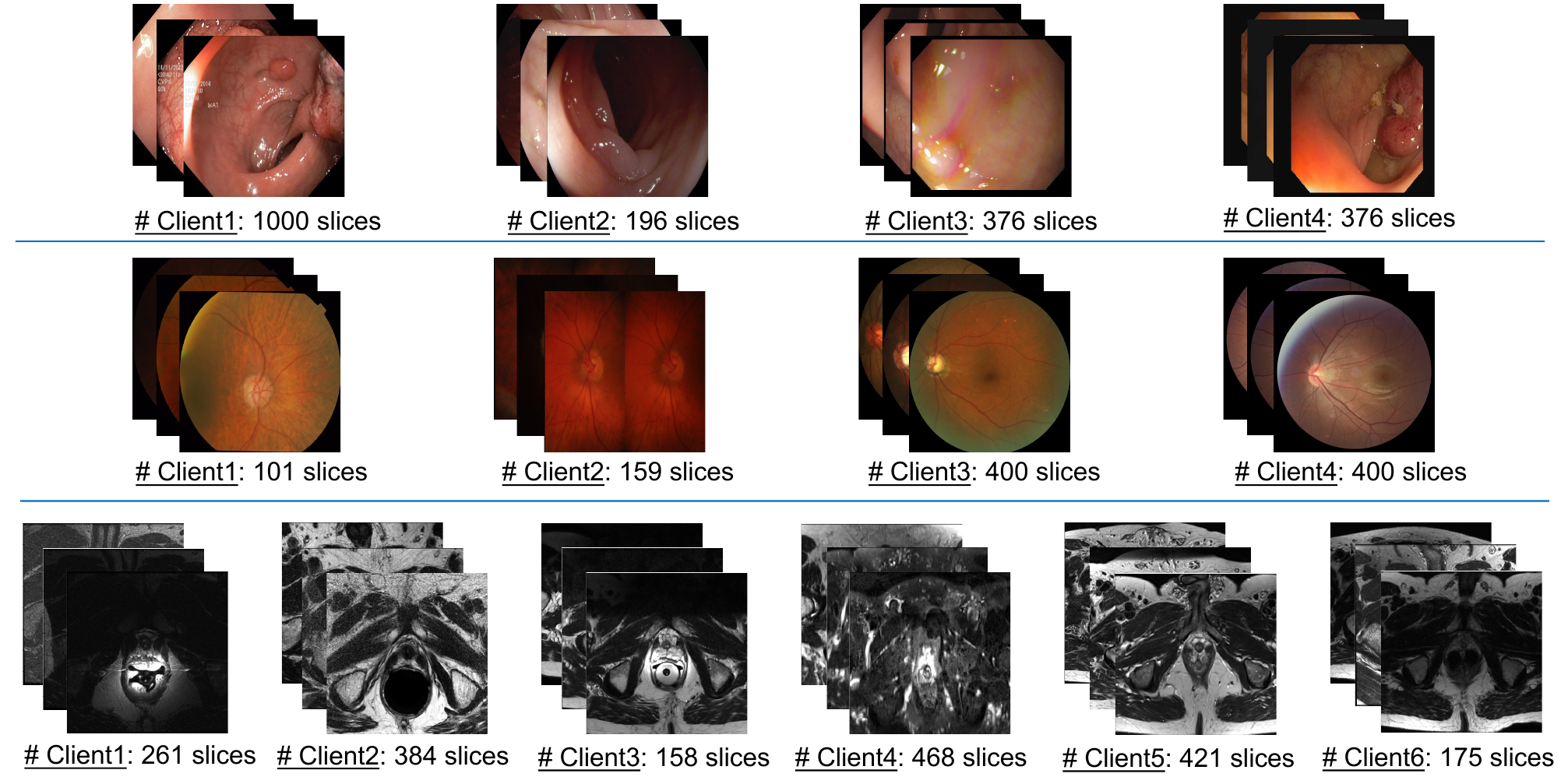}
    \caption{From top to bottom, visualized samples from Fed-Polyp~\cite{wang2022personalizing}, Fed-Fundus~\cite{liu2021feddg}, and Fed-Prostate~\cite{liu2021feddg}, respectively.
    Significant differences in image representation and the sample sizes can be observed between clients.}
    \label{fig:raw_data}
\end{figure}

In our experiments, we followed the setting in~\cite{chen2024think} where each dataset is divided into training and testing at each client with an $8:2$ ratio. 
All raw images of these datasets are randomly cropped to the size of $320 \times 320$ before input~\cite{wu2023medical}.
Visualization of images sampled from these datasets is shown in Figure~\ref{fig:raw_data}.

\subsection{Implementation Details}
This subsection provides a brief view of the implementation details for our experiments.

\vspace{1mm}
\noindent
\textbf{Federated Parameter Efficient Tuning.}
We utilize Medical-SAM~\cite{wu2023medical} to perform medical image segmentation with an efficient adapter tuning approach.
Specifically, we set the training rate to $1e^{-4}$ using the Adam optimizer~\cite{kingma2014adam} at each client.
The batchsize is set to 32, and each client is trained for only one peoch per round.
The total number of rounds, $R$, is set to 100, and $\beta$ is set to 0.01. The federated settings are established using the Flower framework~\cite{beutel2020flower}.

\vspace{1mm}
\noindent
\textbf{Evaluation Metrics.}
We utilize two widely adopted segmentation metrics to thoroughly evaluate our methods: \textbf{IoU}, which represents the intersection-over-union between the prediction and ground truth. 
\textbf{Dice Score}, which indicates the similarity between the predicted pixel set and the annotated pixel set.
All experiments are implemented in PyTorch on a single NVIDIA RTX-3090 GPU.

\vspace{1mm}
\noindent
\textbf{Baseline.} We compare our method with the following methods: 
1) \textbf{FedProx}~\cite{li2020federated} that utilizes the regularization term to reduce the negative effect caused by heterogeneous issues.
2) \textbf{FedFMS (FedAvg)}~\cite{liu2024fedfms} that communicates learnable parameters to the server and then implements average aggregation. 
These two methods are widely recognized baselines in FL and are reproduced in our comparative experiments.
Besides, to comprehensively evaluate our method, we reproduce the following advanced personalized aggregation algorithms: 3) \textbf{FedHEAL}~\cite{chen2024fair} that tackles the domain skew challenge by introducing dynamic local updating and fair global aggregation.
4) \textbf{FedDBE}~\cite{zhang2024eliminating} that addresses the heterogeneous issue by introducing a domain bias eliminator.
Moreover, we also compare the following federated learning algorithms that have been successfully applied to foundation model tuning tasks: 5) \textbf{FedRA}~\cite{su2024fedra} generates a random matrix in each round to guide the aggregation process.
6) \textbf{FedSA}~\cite{guo2024selective} that originally operates on LoRA module that only communicates part of the LoRA, called the global module and remains are preserved in local clients. 
7) \textbf{FedALA}~\cite{zhang2023fedala} that proposed an Adaptive Local Aggregation module (ALA) to aggregate the uploaded models and the global model to achieve the goal of personalized learning in an element-wise manner.

For fairness, we adapt the above methods into our setting where only the adapters inserted in the foundation models are trainable and engaged in the communication process. 

\subsection{Main Experimental Results}
\textbf{Results on Fed-Prostate.}
The experimental results on Fed-Prostate are presented in Table~\ref{tab:tab1}.
Our method demonstrates superior overall performance, achieving a $1.1\sim3.3\%$ improvement in Dice score and a $1.5\sim5.0\%$ increase in IoU.
Notably, our method yields substantial improvement for client $\mathcal{C}_{3}$ compared to the baselines, while consistently enhancing performance across other clients.

\vspace{1mm}
\noindent\textbf{Results on Fed-Polyp.}
We report the experiments on Fed-Polyp in Table~\ref{tab:tab2}, which consists of four clients.
Due to heterogeneous dataset distribution, clients $\mathcal{C}_2$ and $\mathcal{C}_3$ exhibit significant performance drops compared to clients $\mathcal{C}_1$ and $\mathcal{C}_4$.
Compared to FedFMS, which employs naive average aggregation, for client $\mathcal{C}_{2}$, our method achieves $7.0\%$ improvement in Dice score and a $9.1\%$ improvement in IoU.
For client $\mathcal{C}_{3}$, we observe improvements of $8.1\%$ in Dice score and $8.8\%$ in IoU, underscoring the effectiveness of our method in addressing heterogeneity.
On average, our method outperforms the baselines with  $2.0\sim 4.9\%$ gains in Dice score and $1.9\sim 6.2\%$ gains in IoU.

\noindent \textbf{Results on Fed-Fundus.}
For Fed-Fundus, which includes two sub-datasets per client: Optic-Disc and Optic-Cup, we report the average Dice score and IoU across these sub-datasets in Table~\ref{tab:tab2}.
Particularly, compared with baselines, our method obtains $0.2\sim 7.8\%$ improvement in Dice score on average.
For the IoU metric, our method obtains $0.3\sim 10.1\%$ improvement on average.
Notably, while the baselines struggle with performance for client $\mathcal{C}_2$, our method consistently maintains high performance.

\subsection{Empirical Analysis}
In this subsection, we present a series of empirical experiments to demonstrate the superiority of our method.

\setlength{\tabcolsep}{5pt}

\begin{table}[!t]
\caption{Computation and communication overhead on FedPolyp dataset, and the gray background indicates that full-tuning implements the FL algorithms.}
    \centering
    \setlength{\tabcolsep}{1.1mm}
    \begin{tabular}{l|c|c|c}
    \toprule
    \textbf{Methods}  & Result & Computation & Communication \\
    \midrule
    \rowcolor[gray]{0.9} FedAvg-F & 81.4 & 357.4M & 34.9G \\
    \rowcolor[gray]{0.9} FedProx-F & 82.2 & 357.4M & 34.9G\\
    FedAvg & 76.0 & 56.2M & 5.5G\\
    FedProx & 74.9 & 56.2M & 5.5G\\

    FedRA & 67.0 & 56.2M & 3.7G\\ 
    FedSA & 77.1 & 56.2M & 2.8G\\
    Ours & 81.1 & 56.2M & 156.4M\\
    \bottomrule 
    \end{tabular}
    \label{tab:com_compare}
\end{table}


\vspace{1mm}
\noindent \textbf{Computation and Communication Overhead.}
We compare the communication and communication costs on Fed-Polyp and report the IoU performance in Table~\ref{tab:com_compare}.
Specifically, we measure the learnable parameters of one client to represent the computation overhead, and for the communication costs, we summarize the number of parameters engaged in total communication rounds.
The results show that with full-tuning, all foundation model parameters are fully learnable and actively engaged in collaboration, resulting in impractical computational and communication costs while achieving the best segmentation results.
In contrast, our method obtains comparable performance within limited computational resources.
Moreover, compared with other baselines that are implemented on fine-tuning techniques, our method dramatically released the communication cost.


\vspace{1mm}

\begin{table}[!t]
\caption{Ablation experiments (IoU\%) on Fed-Polyp, and the best results are highlighted in \underline{\textbf{black bold}}. $C$ represents overall communication costs.}
    \centering
    \setlength{\tabcolsep}{1.1 mm}{
    \begin{tabular}{l|c c c c c | c}
    \toprule 
    \textbf{Methods} & $\mathcal{C}_1$ & $\mathcal{C}_2$ & $\mathcal{C}_3$ & $\mathcal{C}_4$ & Avg. & $C$\\
    \midrule
    FedAvg & 81.2 & 67.2 & 55.5 & 83.1 & 76.0 & 5.5G \\
    SGCA & 83.0 & 71.3 & 65.8 & 83.5 & 79.1 & 5.5G\\
    LAT & 81.9 & 66.7 & 66.3 & 85.5 & 78.8 & 156.4M \\
    \textbf{Ours} & \underline{\textbf{83.4}} & \underline{\textbf{76.3}} & \underline{\textbf{69.2}} & \underline{\textbf{86.1}} & \underline{\textbf{81.1}} & 156.4M\\
    \bottomrule 
    \end{tabular}}
    \label{tab:abla}
\end{table}

\begin{figure}[!t]
    \centering
    \includegraphics[width=1.0\linewidth]{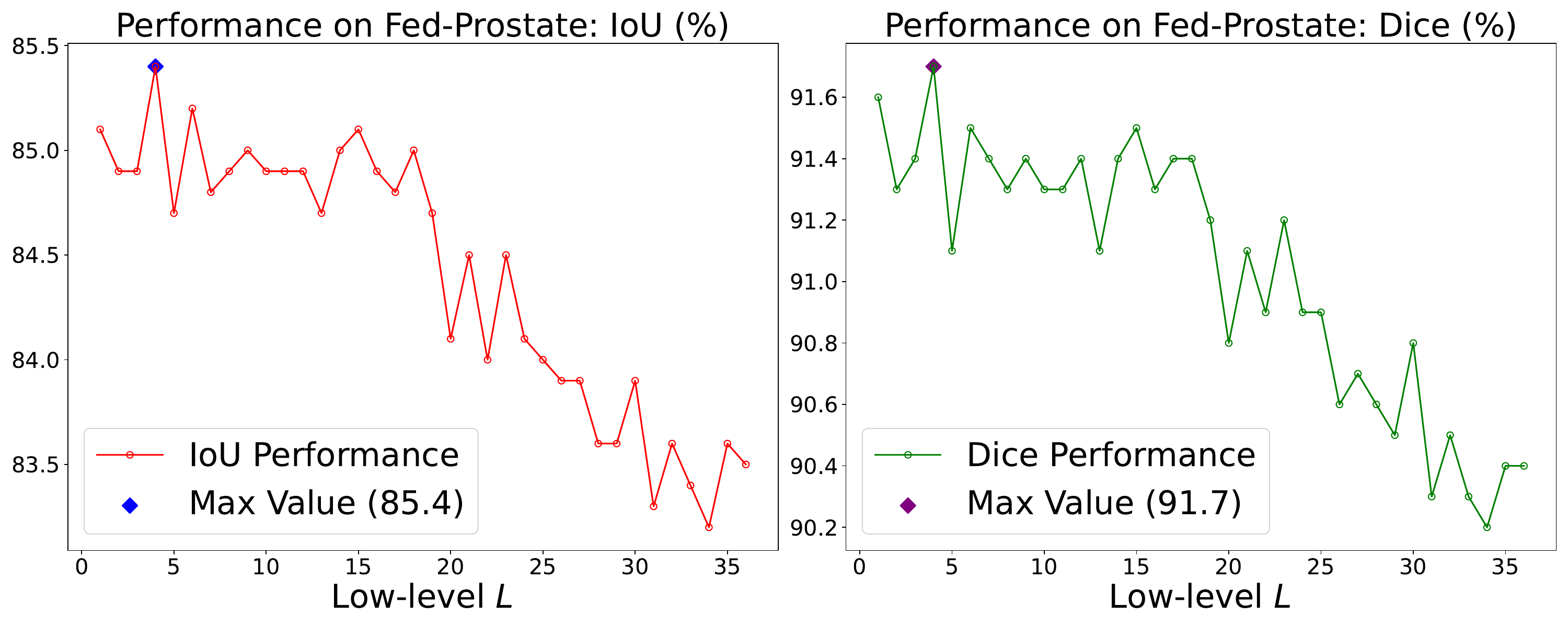}
    \caption{The IoU and Dice change against different $L$ values on the Fed-Prostate dataset.}
    \label{fig:adapters}
\end{figure}
\vspace{1mm}
\noindent \textbf{Ablation Experiments.}
To evaluate the effectiveness of the proposed modules, we conduct ablation experiments on the Fed-Polyp dataset to examine the performance of our method under different scenarios.
Here, we first give the communication cost function:
\begin{equation}
    C = 2 \sum_{r=1}^{R} \sum_{i=1}^{N}|\Theta_{A_{i}}^{(r)}| = 2 \sum_{r=1}^{R}\sum_{i=1}^{N}\sum_{k=1}^{K}|\Theta_{A_i^k}^{(r)}|,
    \label{eq:comm_costs}
\end{equation}
where the total communication round $R$ and the number of clients $N$ are predetermined.
When only communicating $\{\Theta_{A_i^1}\}$, the costs are 156.4M approximately. 
When all $\{A_i^k\}_{k=1}^{K}$ are involved in communication, the costs are 5.5G approximately.
Experimental results and corresponding communication costs are summarized in Table~\ref{tab:abla}.
Specifically, we use FedAvg to denote that all learnable parameters are aggregated using naive average aggregation.
SGCA indicates that all learnable parameters are uploaded to the server and aggregated by using SGCA. 
Besides, LAT denotes that only the learnable parameters $\{\Theta_{A_i^1}\}$ are involved in the average aggregation. 
Compared to FedAvg, SGCA mitigates heterogeneity issues to some extent, yielding significant improvements for clients $\mathcal{C}_2$ and $\mathcal{C}_3$.
Additionally, the LAT results illustrate that the lower layer effectively captures sufficient universal representation knowledge, which significantly enhances local performance while minimizing redundant information.
By leveraging both LAT and SGCA, our method achieves sota performance across all clients within limited communication costs.

\begin{figure}[!t]
    \centering
    \includegraphics[width=1.0\linewidth]{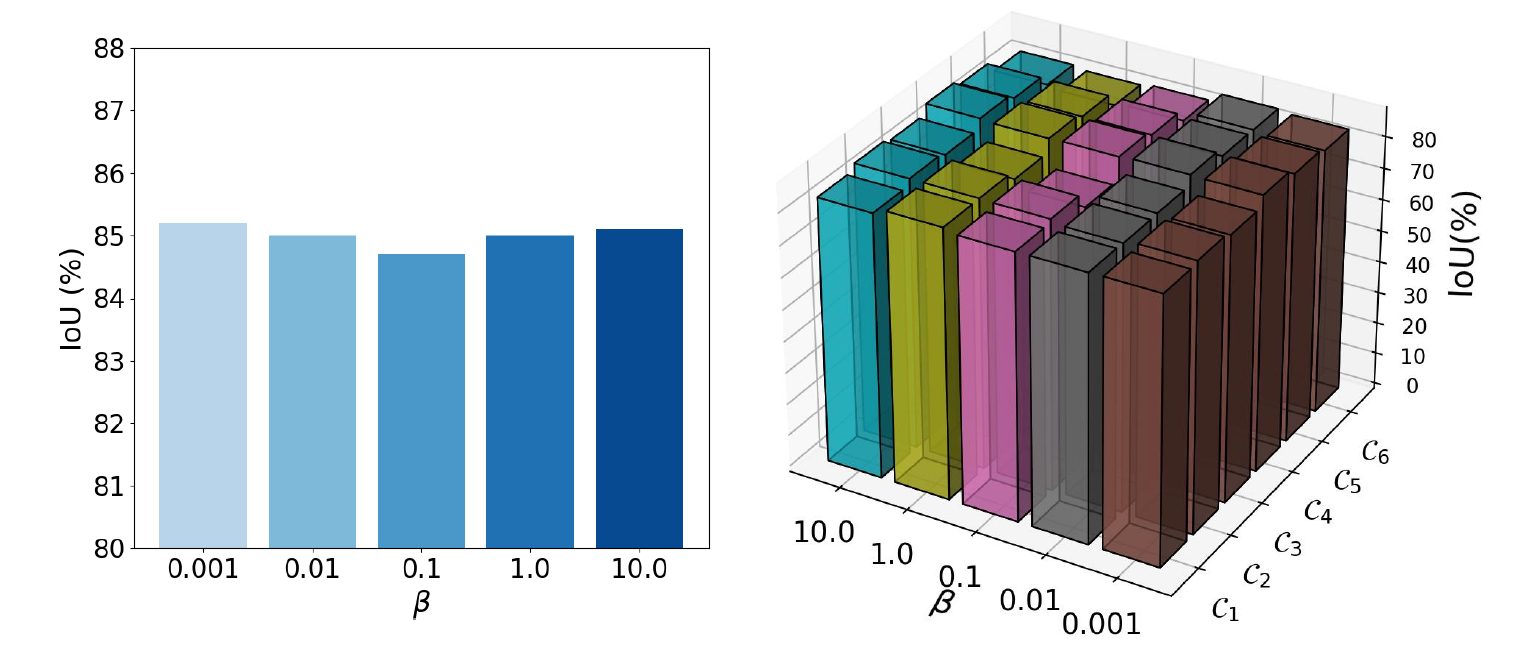}
    \caption{Sensitivity analysis of hyper-parameter $\beta$ on Fed-Prostate dataset.
    We report the IoU performance both on average and for each client.}
    \label{fig:hypersen}
\end{figure}

\vspace{1mm}
\noindent \textbf{Low $L$ Layers Navigation.}
In our main experiments, we set $L=1$ and observed substantial improvements compared to the baselines under identical scenarios. 
To further explore the impact of learnable parameters involved in the communication phase, we varied $L$ from its minimum ($L=1$) to maximum ($L=36$) on the Fed-Prostate dataset.
As shown in Figure~\ref{fig:adapters}, our method reaches the peak at $L=4$, with an IoU of 85.4\%, and the Dice score of 91.7\%.
It can be observed that as the $L$ increases, the final performance has shown a significant downward trend, indicating that not all learnable parameters are essential for collaboration effectiveness.
Instead, excessive parameters may introduce redundant information, impeding performance gains.

\vspace{1mm}
\noindent \textbf{Hyperparameter $\beta$ Sensitive Analysis.}
In our main experiments, we set $\beta=0.01$ in Equation~(\ref{eq:eq9}) to regularize the local training process and effectively alleviate the overfitting issue.
Further, to explore the effect arising from $\beta$, we tune its value within the range $(0.001, 0.01, 0.1, 1.0, 10.0)$ on the Fed-Prostate dataset, and we visualize the results in Figure~\ref{fig:hypersen}.
Specifically, among these six clients, $\mathcal{C}_{3}$ is the most sensitive to changes in $\beta$, achieving its lowest performance ($80.0\%$) at $\beta = 0.1$ and its highest ($84.2\%$ IoU) at $\beta = 10.0$.
In summary, the average performance remains relatively stable across different $\beta$ values, with only a minor $0.5\%$ IoU variation observed.

\vspace{1mm}
\noindent \textbf{Similarity Metric Effect.}
In SGCA, constructing the collaborative relationship matrix is critical for achieving effective and reliable aggregation. 
In our main experiments, we use the $L_2$ norm distance to measure the similarity between clients. 
To further investigate the impact of different similarity measurements, we compare four classic metrics: Cosine similarity, Inner product, $L_1$ norm, and $L_2$ norm, and evaluate their respective performance on the Fed-Polyp datasets.
The results, summarized in Table~\ref{tab:metric}, indicate that the $L_2$-based metric performs best for our method.

\begin{figure}[!t]
    \centering
    \includegraphics[width=1.0\linewidth]{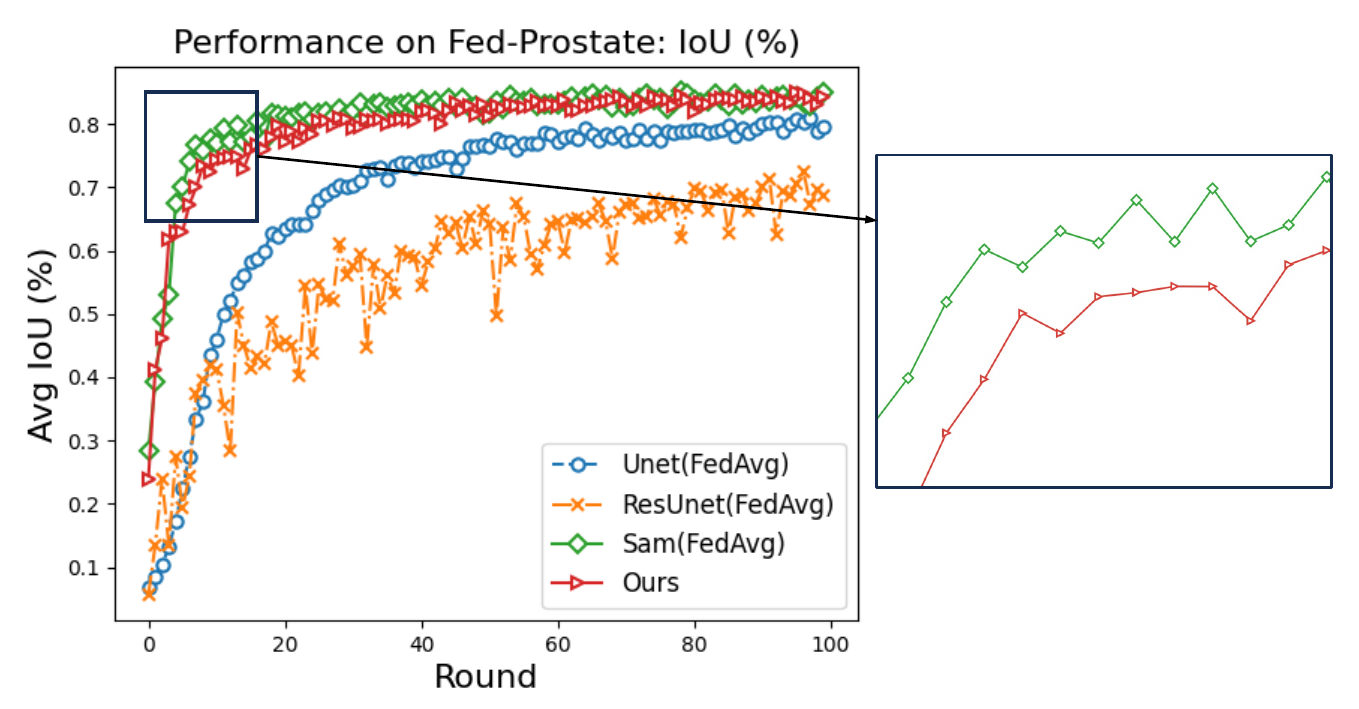}
    \caption{The segmentation performance (IoU\%) against FL rounds with different backbones on the Fed-Prostate dataset.}
    \label{fig:com_backbone}
\end{figure}

\vspace{-1mm}

\begin{table}[!t]
\caption{Comparison results (IoU\%) of different similarity metrics used in SGCA on Fed-Polyp dataset, and the best results are highlighted in \textbf{\underline{black bold}}.}
    \centering
    \begin{tabular}{l|c c c c c c c }
    \toprule 
    \textbf{Metrics} & $\mathcal{C}_1$ & $\mathcal{C}_2$ & $\mathcal{C}_3$ & $\mathcal{C}_4$ &  Avg. \\
    \midrule
    Inner & 83.9  & 75.3 & 63.8 & 86.2 & 80.3 \\

    Cosine & 83.7 & \underline{\textbf{78.4}} & 62.3 & \underline{\textbf{86.7}} & 80.4  \\
    $L_1$-based & \underline{\textbf{84.1}} & 69.1 & 66.5 & 86.4 & 80.4 \\
    $L_2$-based & 83.4 & 76.3 & \underline{\textbf{69.2}} & 86.1 & \underline{\textbf{81.1}} \\
    \bottomrule 
    \end{tabular}
    \label{tab:metric}
\end{table}
\vspace{-1mm}

\noindent \textbf{Comparison with Convolution-based Backbone.}
To demonstrate the impact of foundation models in facilitating FL aggregation convergence, we compare our results with classical convolution-based backbones, including UNet~\cite{ronneberger2015u}, ResUnet~\cite{zhang2018road}, Sam (full-tuning), and our method. 
As shown in Figure~\ref{fig:com_backbone},  compared to UNet, the ResUnet architecture includes additional layers and residual connections, increasing model complexity and making convergence more challenging.
Benefiting the powerful capacity of foundation models, Sam (full-tuning) attains the highest performance within the fewest rounds. 
Our method strikes an optimal balance, achieving competitive performance while significantly reducing resource demands.

\section{Conclusion}

In this paper, we proposed a novel FLFM fine-tuning method, named FedSCA, for the medical image segmentation task, enabling collaborative model training across multiple hospitals while preserving patient privacy.
Our proposed method leverages the combined strength of FL and FM, effectively addressing non-IID challenges while simultaneously minimizing computational and communication demands. 
Specifically, we leverage PEFT to enable each client to adapt the deployed foundation segmentation model within manageable computational limits, maintaining practical resource usage. 
Recognizing that not all learnable parameters are necessary for sharing generalizable representation knowledge, we transmit only the learnable parameters from low-level adapters, thereby reducing redundant information and significantly lowering communication costs. 
Additionally, to further mitigate non-IID issues, we developed SGCA, which enables clients with similar data distributions to benefit more substantially from each other, reducing the negative impacts associated with data heterogeneity. 
Extensive experiments demonstrate the effectiveness and efficiency of FedSCA, achieving a new SOTA performance while substantially reducing communication costs. 

\bibliography{example_paper}
\bibliographystyle{icml2025}


\end{document}